\documentclass[journal,letterpaper, preprint]{IEEEtran}

\usepackage[dvips]{graphicx} 
\usepackage{amsfonts}
\usepackage{amscd}
\usepackage{amsmath}    

\begin{document}

\title{Unconditional security at a low cost}
\author{Xiongfeng Ma \\
Center for Quantum Information and Quantum Control,\\
Department of Physics, University of Toronto, Toronto, Ontario, Canada\\
}

\maketitle

\begin{abstract}
By simulating four quantum key distribution (QKD) experiments and
analyzing one decoy-state QKD experiment, we compare two data
post-processing schemes based on security against individual attack
by L\"{u}tkenhaus, and unconditional security analysis by
Gottesman-Lo-L\"{u}tkenhaus-Preskill. Our results show that these
two schemes yield close performances. Since the Holy Grail of QKD is
its unconditional security, we conclude that one is better off
considering unconditional security, rather than restricting to
individual attacks.
\end{abstract}

\section{Introduction}
Quantum key distribution (QKD) \cite{BB_84,Ekert_91} allows two
parties, a transmitter Alice and a receiver Bob, to create a random
secret key even when the channel is accessible to an eavesdropper,
Eve. The security of QKD is built on the fundamental laws of physics
in contrast to existing classical public key encryption schemes that
are based on unproven computational assumptions. The unconditional
security of the idealized QKD system has been proven in the past
decade \cite{Mayers_01,LoChauQKD_99,ShorPreskill_00,GRTZ_02}.

In Shor-Preskill's proof \cite{ShorPreskill_00}, two steps of data
post-processing, error correction and privacy amplification, need to
be performed in order to distill a secret key. Error correction
ensures that Alice and Bob share an identical key, while privacy
amplification removes Eve's information about the final key. Alice
and Bob can simply estimate the quantum bit error rate (QBER) by
error testing and then perform error correction. Privacy
amplification, on the other hand, requires the phase error rate
which cannot be measured directly without a quantum computer. When
an idealized QKD system is used, due to the symmetry of BB84, one
can assume the phase error rate to be the same as the QBER
\cite{ShorPreskill_00}. However, any real QKD setup is not ideal but
with imperfect sources, noisy channels and inefficient detectors,
which will affect the security. To do security analysis, we should
take these effects into consideration.

A few theoretical works have been done to deal with the imperfect
devices, such as
\cite{MayersYao_98,IndividualAttack_00,BLMS_00,FGSZ_01,ILM_01,KoashiPreskill_03,GLLP_04}.
We will compare L\"{u}tkenhaus' analysis \cite{IndividualAttack_00}
which deals with individual attacks and
Gottesman-Lo-L\"{u}tkenhaus-Preskill (GLLP) \cite{GLLP_04}
unconditional security proof. For convenience, we name the data
post-procesing schemes, based on these two security analyses, after
L\"{u}tkenhaus and GLLP.


Meanwhile, many QKD experiments
\cite{BBBSS_92,Townsend_98,RGGGZ_98,BGKHJTLS_99,GYS_04} have been
performed in the past decade. Experimentalists sometimes use the
QBER as the only criterion for the security of QKD.
However, after taking the imperfections into consideration, this
kind of security analysis is incomplete. In fact, due to
photon-number splitting (PNS) attacks
\cite{HIGM_95,BLMS_00,LutkenhausJahma_02}, Eve can successfully
break down the security even when the QBER is 0\%.

Decoy states have been proposed as a useful method for substantially
improving the performance of QKD with coherent sources
\cite{Hwang_03}. The security proof of decoy-state QKD is given in
Ref.~\cite{Decoy_05}. Afterwards, some practical decoy-state
protocols are proposed \cite{Wang_05,HEHN_05,Practical_05}.
Recently, a few decoy-state QKD experiments have been performed
\cite{ZQMKQ_06,ZQMKQ60km_06,PanDecoy_06,LosAlamosTES_06}. In this
paper, we will consider both decoy-state and non-decoy-state cases.



The goal of this paper is to compare the two standard security proof
results---L\"utkenhaus and GLLP. We find that for realistic
experimental parameters, with or without decoy states, the two
security proof results give similar key generation rate and secure
distance. Since unconditional security is the Holy Grail of QKD and
GLLP (but not L\"utkenhaus) gives unconditional security, our
conclusion is that one should use GLLP as the standard criterion for
security.



In Section \ref{Model} we review a widely used QKD model to real
experiments. In Section \ref{PostPro}, we will compare two data
post-processing schems, L\"utkenhaus and GLLP for non-decoy and
decoy-state QKD.


\section{Preliminaries} \label{Model}
Here we use a QKD model following \cite{IndividualAttack_00}, see
also \cite{Practical_05}. We do not repeat the details of the model
here. To reproduce the simulation results, one may need to refer to
Section II of \cite{Practical_05}.

The procedure of a QKD experiment, using BB84 protocol with coherent
state, is as follows:
\begin{enumerate}
\item In total, Alice sends Bob $N$ pulses for QKD,
containing $N_\mu$ pulses for key transmission (signal states) and
$N-N_\mu$ pulses for error testing or decoy states. In the $N_\mu$
signal pulses, Alice and Bob have $N_\mu^s$ pulses using same bases
(after basis reconciliation).
\item Within the $N_\mu^s$ pulses, Bob gets a sifted key with a length of
$K_\mu^s$, where they measure the same bases and Bob gets
detections.
\item Alice and Bob choose a security analysis and perform a data post-processing scheme.
\end{enumerate}


Here we assume Alice uses a weak coherent state for key
transmission.
Define the expected photon number (intensity) of the weak coherent
state as $\mu$.

Define $q=N_\mu^s/N$. In BB84 scheme, $q=1/2$ when
$N\rightarrow\infty$. Here subscript $\mu$ is the expected photon
number of the coherent light used for key transmission as defined
above.

Define the gain $Q_\mu=K_\mu^s/N_\mu^s$, the probability for Bob to
get a detection in a pulse that Alice and Bob use the same basis.

Define the QBER $E_\mu=K_\mu^{err}/K_\mu^s$, the probability for Bob
to get a wrong detection in a pulse that Alice and Bob use the same
basis. Here $K_\mu^{err}$ is the number of erroneous bits in the
sifted key.


Alice knows what $N$ and $\mu$ she uses for the key transmission.
$N_\mu^s$ and $K_\mu^s$ can be directly counted from the data after
the key transmission. Alice and Bob can estimate QBER from error
testing, or they can count $K_\mu^{err}$ after error correction. In
fact, even without knowing the real QBER, they can directly apply a
error correction scheme (e.g., the Cascade scheme
\cite{BrassardSalvail_93}). If the error correction is successful,
then it automatically provides the QBER, otherwise they restart the
QKD. Thus, in a real QKD system, Alice and Bob may skip the error
testing part.

Assuming that the phase of each pulse is totally randomized, the
photon number of each pulse follows a Poisson distribution with a
parameter $\mu$ as its expected photon number. We remark that the
phase randomization procedure is crucial for the security of QKD
\cite{LoPreskill_05}. The density matrix of the state emitted by
Alice is given by
\begin{equation}\label{Model:AliceState}
\rho_A=\sum^{\infty}_{i=0}\frac{\mu^i}{i!}\,e^{-\mu}\,
|i\rangle\langle i|,
\end{equation}
where $|0\rangle\langle 0|$ is the \textit{vacuum state} and
$|i\rangle\langle i|$ is the $i$-photon state for $i=1,2\cdots$. The
states with only one photon ($i=1$) are called \emph{single photon
states}. The states with more than one photon ($i\ge2$) are called
\textit{multi photon states}. 

Define $Y_i$ to be the \emph{yield} of an $i$-photon state, i.e.,
the conditional probability of a detection event at Bob's detector
given that Alice sends out an $i$-photon state. Note that $Y_0$ is
the background rate including detector dark counts and other
background contributions such as the stray light in the fiber.
Consequently, define the \emph{error rate} of $i$-photon state to be
$e_i$. The \emph{gain} of $i$-photon states $Q_i$ is given by
\begin{equation}\label{Model:Qi}
\begin{aligned}
Q_i &= Y_i\frac{\mu^i}{i!}e^{-\mu}.
\end{aligned}
\end{equation}
When $i=1$, $Q_1$ and $e_1$ are the gain and error rate of single
photon states. Note that Eve has the ability to change $\{Y_i\}$ and
$\{e_i\}$ as she wishes, but she cannot change $\mu$, which is set
by Alice. Decoy states allow Alice and Bob to estimate channel
transmittance and error rate accurately, which will restrict the
Eve's freedom to adjust $\{Y_i\}$ and $\{e_i\}$. This is the key
reason why decoy states are useful for QKD \cite{Decoy_05}.

\section{Data post-processing schemes} \label{PostPro}
In this section, we will compare two data post-processing schemes,
L\"utkenhaus versus GLLP. During the comparison, we will apply two
data post-processing schemes to both non-decoy and decoy state QKD.

\subsection{L\"utkenhaus versus GLLP} \label{Compare}
Here we compare data post-processing schemes based on two security
analyses, L\"utkenhaus \cite{IndividualAttack_00} and GLLP
\cite{GLLP_04}. L\"utkenhaus scheme focuses on security against
individual attacks, while GLLP scheme proves unconditional security.

In the GLLP \cite{GLLP_04}, there are so called tagged qubits in the
discussion. The basis information of tagged qubits is somehow
revealed to Eve. Thus tagged qubits are insecure for QKD. The idea
is that Alice and Bob can (in principle) separate the qubits into
two groups, tagged and untagged qubits, hence they only need to
perform privacy amplification to the untagged qubits. The reason is
as follows. The final key will be bitwise $XOR$ of keys that could
be obtained from the tagged and untagged qubits. If the key from
untagged qubits is private and random, then it doesn't matter if Eve
knows everything about tagged qubits --- the sum is still private
and random.

Based on the tagged qubit idea, the procedure of the data
post-processing is as follows. First, Alice and Bob perform the
error correction, and if it is successfully done, they share an
identical key. And then they calculate how much privacy
amplification they should do (according to certain security
analysis, which we will discuss soon). Finally they use random
hashing (or other privacy amplification procedure) to get a final
secure key.

Due to PNS attack, we can regard qubits single photon states as
untagged qubits and those from other (vacuum and multi photon)
states as tagged qubits. Then we can compare the results of
L\"{u}tkenhaus and GLLP schemes.  We can rewrite the formula of the
key generation rate by L\"{u}tkenhaus scheme
\begin{equation}\label{TwoSch:Lutkenhaus}
\begin{aligned}
R\geq q\{-Q_\mu H_2(E_\mu)+Q_1[1-\log_2(1+4e_1-4e_1^2)]\}.
\end{aligned}
\end{equation}
Similarly, as given in Eq.~(11) in \cite{Decoy_05}, we rewrite the
key rate formula of GLLP
\begin{equation}\label{TwoSch:GLLP}
\begin{aligned}
R\geq q\{-Q_\mu H_2(E_\mu)+Q_1[1-H_2(e_1)]\}
\end{aligned}
\end{equation}
where $Q_1$ and $e_1$ are the gain and error rate of single photon
states, and $H_2(x)=-x\log_2(x)-(1-x)\log_2(1-x)$ is the binary
entropy function. Alice and Bob need to estimate $Q_1$ and $e_1$
given the data from QKD experiments.


In both Eqs.~\eqref{TwoSch:Lutkenhaus} and \eqref{TwoSch:GLLP}, the
first term in the bracket is for error correction and the second one
is for privacy amplification. The privacy amplification is only
performed on the single photon part.  In this manner, L\"utkenhaus
\cite{IndividualAttack_00} has already applied the tagged qubits
idea.

Here due to PNS attacks \cite{HIGM_95,BLMS_00,LutkenhausJahma_02},
both L\"utkenhaus and GLLP assume that the single photon states are
the only source of untagged qubits for BB84. This may not true for
other protocols. For example, in SARG04 \cite{SARG_04,TamakiLo_06},
two-photon states can be used to extract secure keys.

\begin{figure}[hbt]
\centering \resizebox{8cm}{!}{\includegraphics{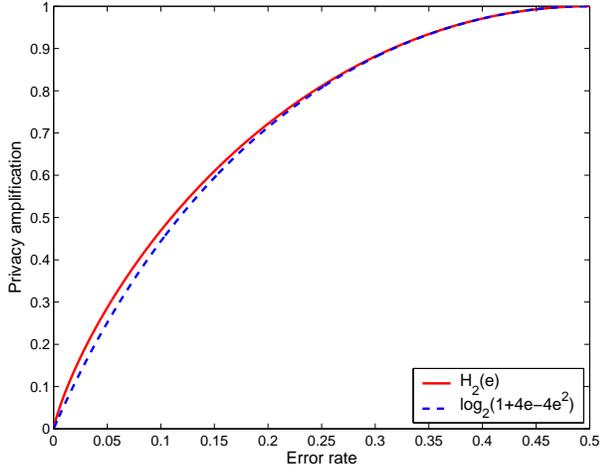}}
\caption{GLLP vs.~L\"utkenhaus. The maximal deviation of two curves
is 15.36\% when the error rate is 3.85\%.} \label{TwoSch:fig:ComPri}
\end{figure}

Here is the key point of the whole paper. The difference between the
L\"utkenhaus and GLLP results appears in the privacy amplification
part. We compare $H_2(e)$ with $\log_2(1+4e_1-4e_1^2)$ in
Fig.~\ref{TwoSch:fig:ComPri}. We can see that the difference of two
functions are quite small. For this reason, in fact, L\"utkenhaus
and GLLP give very similar result in the key generation rate and
distance of secure QKD. In what follows, we will illustrate this
crucial point with examples of experimental parameters from previous
QKD experiments. Our conclusion holds with and without using decoy
states.

\subsection{Non-decoy-state QKD} \label{nondecoy}
Without decoy states, Alice and Bob have to pessimistically assume
that all losses and errors come from single photon states. Thus
\begin{equation}\label{TwoSch:PessAssum}
\begin{aligned}
Q_1 &= Q_\mu-p_M \\
e_1 &= \frac{Q_\mu E_\mu}{Q_1}
\end{aligned}
\end{equation}
where $p_M=1-(1+\mu)\exp(-\mu)$ is the probability that Alice sends
out a multi photon state. We can recover Eq.~(15) in
\cite{IndividualAttack_00} by substituting
Eq.~\eqref{TwoSch:PessAssum} into Eq.~\eqref{TwoSch:Lutkenhaus}. Let
$\Delta=p_M/Q_\mu$, we can recover Eq.~(50) in \cite{GLLP_04} by
substituting Eq.~\eqref{TwoSch:PessAssum} into
Eq.~\eqref{TwoSch:GLLP}.

We compare the key generation rate of two data post-processing
schemes based on L\"utkenhaus and GLLP by simulating four experiment
setups \cite{Townsend_98,RGGGZ_98,BGKHJTLS_99,GYS_04}. The key
parameters are listed in TABLE \ref{TwoSch:Table:exdata}.

\begin{table}[h]\center
\begin{tabular}{|c|c|c|c|c|}
\hline
& T8\cite{Townsend_98} & G13\cite{RGGGZ_98} & KTH\cite{BGKHJTLS_99} & GYS\cite{GYS_04}\\
\hline
$\lambda$ [nm] & 830 & 1300 & 1550 & 1550 \\
\hline
$\alpha$ [dB/km] & 2.5 & 0.32 & 0.2 & 0.21 \\
\hline
$e_{d}$ [\%]& 1 & 0.14 & 1 & 3.3 \\
\hline
$Y_0$ [/pulse] & $10^{-7}$ & $1.64\times10^{-4}$ & $4\times10^{-4}$ & $1.7\times10^{-6}$ \\
\hline
$\eta_{Bob}$ [\%]& 7.92 & 8.14 & 14.30 & 4.5 \\
\hline
\end{tabular}
\caption{\normalfont{Key parameters from four QKD experiment
setups.}} \label{TwoSch:Table:exdata}
\end{table}

\begin{figure}[hbt]
\centering \resizebox{8cm}{!}{\includegraphics{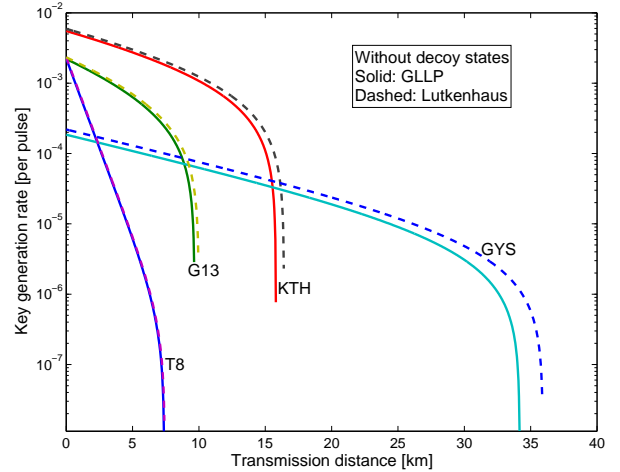}}
\caption{shows the relationship between key generation rate and the
transmission distance, comparing two data post-processing schemes
based on L\"{u}tkenhaus and GLLP security analyses. The key
parameters are listed in TABLE~\ref{TwoSch:Table:exdata}. Here we
consider non-decoy case and use the optimal expected photon number
$\mu=\eta$ \cite{IndividualAttack_00}. Using Cascade protocol
\cite{BrassardSalvail_93}, the error correction efficiency is 1.16.
Details of the QKD simulation model appear in \cite{Practical_05}.}
\label{TwoSch:fig:LutGLLP}
\end{figure}

Fig.~\ref{TwoSch:fig:LutGLLP} shows the relationship between key
generation rate and the transmission distance, comparing two data
post-processing schemes, L\"{u}tkenhaus and GLLP. For both schemes,
we consider non-decoy state QKD. From Fig.~\ref{TwoSch:fig:LutGLLP},
we can see that the key generation rate of GLLP is only slightly
lower than that of L\"{u}tkenhaus. Here we emphasize that GLLP deals
with the general attack, while L\"{u}tkenhaus is restricted to
individual attack.

We remark that the QBERs ($E_\mu$'s) of four GLLP curves at maximal
distances in Fig.~\ref{TwoSch:fig:LutGLLP} are $4.57\%$, $4.80\%$,
$4.80\%$ and $4.34\%$. Clearly it is far away from $11\%$, the
tolerable QBER given in \cite{ShorPreskill_00}. This is due to the
fact that the QKD source is weak coherent state, while the security
proof given in \cite{ShorPreskill_00} is based on single photon
source.

Many of QKD experiments used $\mu=0.1$ as for key transmission. If
using $\mu=0.1$ for GYS \cite{GYS_04} setup, we find that
$Q_\mu<p_M$ for all transmission distances. That is, for BB84, Eve
can successfully perform PNS attacks
\cite{HIGM_95,BLMS_00,LutkenhausJahma_02} and obtain all the
information about the key even when the QBER is 0\%!


\subsection{Decoy-state QKD} \label{Decoy}
Here, we will give a data post-processing scheme following the
security analysis of decoy state QKD \cite{Decoy_05}.

In Eqs.~\eqref{TwoSch:Lutkenhaus} and \eqref{TwoSch:GLLP}, $q$ and
$Q_\mu$ can be directly counted from Bob's detection events. The
QBER $E_\mu$ can be obtained from error testing or after error
correction step. Alice and Bob can estimate $Q_1$ and $e_1$ with
decoy states. Definitions of these variables are in Section
\ref{Model}.

As for Vacuum+Weak decoy state scheme \cite{Practical_05}, besides
$Q_\mu$ and $E_\mu$ discussed in Section \ref{Model}, Alice and Bob
will use $Q_{vac}$ (from the vacuum decoy), $Q_\nu$ and $E_\nu$
(from the weak decoy). The definitions are similar as $Q_\mu$ and
$E_\mu$ in Section \ref{Model}. The intensity of weak decoy state is
$\nu$. Alice and Bob can publicly compare all weak decoy states to
get $Q_\nu$ and $E_\nu$. They can estimate the background count rate
by vacuum decoy states $Y_0=Q_{vac}$. Then, they apply the formulas,
Eq.~(35) and Eq.~(37) in \cite{Practical_05}, for the estimations of
$Q_1$ and $e_1$
\begin{equation}\label{TwoSch:V+W}
\begin{aligned}
Q_1 &\ge \frac{\mu^2e^{-\mu}}{\mu\nu-\nu^2}(Q_\nu e^{\nu}-Q_\mu
e^\mu\frac{\nu^2}{\mu^2}-\frac{\mu^2-\nu^2}{\mu^2}Y_0) \\
e_1 &\le \frac{E_\nu Q_\nu e^{\nu}-e_0Y_0}{Q_1e^{\mu}\nu/\mu},
\end{aligned}
\end{equation}
where $e_0=1/2$ is the error rate of vacuum decoy states.

In summary, the data post-processing for decoy state QKD is
\begin{enumerate}
\item Alice announces to Bob which pulses are used for decoy states.
They publicly compare all values of decoy states, and then calculate
$Q_{vac}$, $Q_\nu$ and $E_\nu$.
\item They sacrifice $f(E_\mu) H(E_\mu)$ part of the sifted key to do the
error correction. Here $f(E_\mu)$ is the error correction
efficiency.
\item Alice and Bob estimate the gain $Q_1$ and error rate $e_1$
of single photon states, using Eq.~\eqref{TwoSch:V+W} with
$Y_0=Q_{vac}$, $Q_\nu$ and $E_\nu$.
\item They calculate the final key rate $R$ by Eq.~\eqref{TwoSch:Lutkenhaus} or
\eqref{TwoSch:GLLP}. According to $R$, they perform privacy
amplification (say, random hashing) to get a final secure key with
length of $NR$, where $N$ the total number of pulses sent by Alice
defined in Section \ref{Model}.
\end{enumerate}

We remark that in principle Alice and Bob can use decoy states to
generate keys, but in practice it is more efficient if Alice and Bob
compare all bit values of decoy states to minimize the statistical
fluctuation. In other words, there always exists one optimal
intensity for QKD and we use it for signal states. Suppose some of
the decoy state pulses are used for key generation; it will be more
efficient to transmit these pulses using the signal (optimal)
intensity. Thus, only the signal pulses are used for key generation.

Note that for a finite length QKD, Alice and Bob need to consider
statistical fluctuations. A similar procedure will still be
applicable. The only difference will be the formulas for estimations
of $Q_1$ and $e_1$. Statistical fluctuations are discussed in
\cite{Wang_05,Practical_05}. As mentioned in \cite{Practical_05}, a
rough way to estimate the statistical fluctuations is assuming
Gaussian distribution of $Q_\nu$, $E_\nu$ and $Y_0$. Take the lower
bound of $Q_\nu$ and $Y_0$ and the upper bound of $E_\nu$ to
estimate $Q_1$ and $e_1$. Other procedures are the same as used
above. For simplicity, here we skip the statistical fluctuations.

Note that the efficiency of error correction and privacy
amplification can also be included into
Eq.~\eqref{TwoSch:Lutkenhaus} and \eqref{TwoSch:GLLP}. In our
simulations, we only consider the efficiency of error correction.

\begin{figure}[hbt]
\centering \resizebox{8cm}{!}{\includegraphics{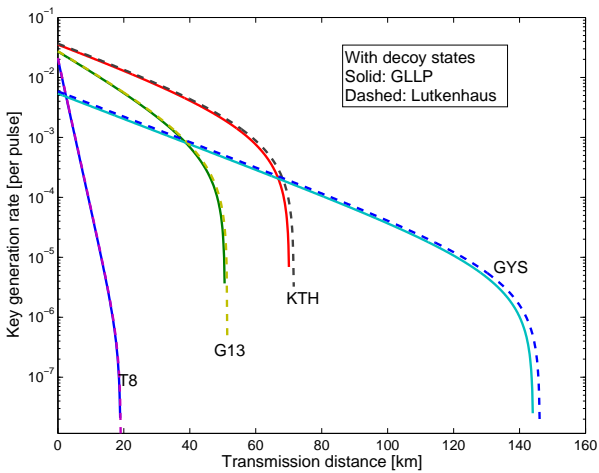}}
\caption{shows the relationship between key generation rate and the
transmission distance for decoy state QKD, comparing L\"{u}tkenhaus
and GLLP. The key parameters are listed in
TABLE~\ref{TwoSch:Table:exdata}. Using the Cascade protocol
\cite{BrassardSalvail_93}, the error correction efficiency is 1.16.
Here we assume the efficiency of privacy amplification is 1. Details
of QKD simulations can be seen in \cite{Practical_05}.}
\label{TwoSch:fig:decoy}
\end{figure}

The comparison of L\"{u}tkenhaus and GLLP for decoy-state QKD is
shown in Fig.~\ref{TwoSch:fig:decoy}. From the figure, we can see
that the performance of two schemes are very close when decoy states
are used.

For comparison, we list the QBERs of four GLLP curves at maximal
distance in Fig.~\ref{TwoSch:fig:decoy} are $5.19\%$, $4.21\%$,
$5.11\%$ and $6.8\%$. We can see that these four values are close to
those given by Fig.~\ref{TwoSch:fig:LutGLLP} of non-decoy state QKD.
This is because stronger signals are allowed to use when decoy
states are implemented, and then the QBER drops down, which cancels
out the increase of QBER by higher channel loss.

Based on the simulation results of four QKD setups, we find that
there is little to gain by restricting the security analysis to
individual attacks, given that the two schemes---L\"{u}tkenhaus
vs.~GLLP---provide very close performances. In other words, our view
is that one is better off considering unconditional security, rather
than restricting to individual attacks.


\subsection{One example} \label{DecoyExp}
Decoy state QKD experiments has recently been performed
\cite{ZQMKQ_06,ZQMKQ60km_06,PanDecoy_06,LosAlamosTES_06}. We analyze
a decoy state QKD experiment over 60km fiber experiment
\cite{ZQMKQ60km_06} as an example here. All raw experiment data are
listed in TABLE~\ref{TwoSch:Table:Ourdata}.

\begin{table}[h]\center
\begin{tabular}{|c|c|c|c|c|c|c|c|c|c|c|c|}
\hline
Distance & $\lambda$ & $N$ & $N_{vac}$ & $K_{vac}$  \\
\hline
$60km$ & $1550nm$ & $104.8Mb$ & $16.63Mb$ & $1033b$ \\
\hline
$\mu$ & $N_\mu$ & $N_\mu^s$ & $K_\mu^s$ & $K_\mu^{err}$ \\
\hline
$0.55$ & $66.86Mb$ & $33.40Mb$ & $60.50kb$ & $1845b$ \\
\hline
$\nu$ & $N_\nu$ & $N_\nu^s$ & $K_\nu^s$ & $K_\nu^{err}$ \\
\hline
$0.152$ & $21.34Mb$ & $10.69Mb$ & $5397b$ & $455b$ \\
\hline
\end{tabular}
\caption{\normalfont{Raw QKD experiment parameters from
\cite{ZQMKQ60km_06}. The unit $b$ stands for bit.}}
\label{TwoSch:Table:Ourdata}
\end{table}

From TABLE~\ref{TwoSch:Table:Ourdata}, we can calculate the key
parameters for security analysis, listed in
TABLE~\ref{TwoSch:Table:Para}. The definitions are given in Section
\ref{Model}.
\begin{table}[h]\center
\begin{tabular}{|c|c|c|c|c|c|c|c|c|c|c|c|}
\hline $q$ & $Q_\mu$ & $E_\mu$ & $Y_0=Q_{vac}$ & $Q_\nu$ \\ 
\hline
$0.319$ & $1.81\times10^{-3}$ & $3.05\%$ & $1.11\times10^{-4}$ & $5.47\times10^{-4}$ \\ 
\hline
\end{tabular}
\caption{\normalfont{Key parameters for the security analysis of
\cite{ZQMKQ60km_06} derived from TABLE~\ref{TwoSch:Table:Ourdata}.}}
\label{TwoSch:Table:Para}
\end{table}

Now, we can apply the data post-processing for decoy state QKD. We
can estimate $Q_1$ by Eq.~\eqref{TwoSch:V+W}. For $e_1$ we use a
different formula
\begin{equation}\label{TwoSch:e1}
\begin{aligned}
e_1 &\le \frac{E_\mu Q_\mu e^{\mu}-e_0Y_0}{Q_1e^{\mu}}.
\end{aligned}
\end{equation}
The reason is that in the real experiment \cite{ZQMKQ60km_06}, $e_1$
of decoy states deviates largely from that of signal states. The
deviation is caused by the imperfections of attenuators. Thus, we
use the QBER of signal states $E_\mu$ to estimate $e_1$. It reminds
us the key assumption of decoy state QKD: all $Y_i$ and $e_i$ are
the same in the signal states and in the decoy states
\cite{Decoy_05}. 

Substituting parameters of TABLE~\ref{TwoSch:Table:Para} into
Eqs.~\eqref{TwoSch:V+W} and \eqref{TwoSch:e1}, we get
$Q_1\ge8.50\times10^{-4}$ and $e_1\le2.73\%$. Then we apply the
Cascade error correction scheme \cite{BrassardSalvail_93},
sacrificing a fraction of $1.16 H_2(E_\mu)=0.486$ of the sifted key,
where $1.16$ is the error correction efficiency. Then from
Eqs.~\eqref{TwoSch:Lutkenhaus} and \eqref{TwoSch:GLLP}, we get the
key generation rate $R_{Lutkenhaus}=9.98\times10^{-5}$ and
$R_{GLLP}=9.04\times10^{-5}$ . We randomly choose parities and
obtain a final key with length of
$K_{Lutkenhaus}=NR_{Lutkenhaus}=10.5kbit$ and
$K_{GLLP}=NR_{GLLP}=9.47kbit$ . We can see the difference between
the key lengths of GLLP and L\"utkenhaus is within 10\%.
For the case of considering statistical fluctuations, one can refer
to \cite{ZQMKQ60km_06}.

%
%

\section{Conclusion}
In this paper, we compare the security analysis of L\"{u}tkenhaus,
against individual attack, and Gottesman-Lo-L\"{u}tkenhaus-Preskill,
general security analysis. Our simulation results show that these
two schemes provide close performances. Thus, we conclude that one
is better off considering unconditional security, rather than
restricting to individual attacks. In the security analysis, we
emphasize that the QBER is not the only criterion of security due to
the imperfections of QKD setups. 

\section{Acknowledgments}
This work is mostly from Xiongfeng Ma's Master Thesis (see ArXiv:
quant-ph/0503057) under the supervision of Hoi-Kwong Lo at
University of Toronto. We thank Chi-Hang Fred Fung,
N.~L\"{u}tkenhaus, Bing Qi and Yi Zhao for enlightening discussions.
Financial support from University of Toronto, CFI, CIAR, CIPI,
Connaught, CRC, NSERC, OIT, PREA and Chinese Government Award for
Outstanding Self-financed Students Abroad is gratefully
acknowledged.

\bibliographystyle{ieeetr}

\bibliography{Bibli}


\end{document}